\documentclass{revtex4}

\usepackage{graphicx}
\usepackage{amssymb}
\usepackage{amsmath}
\usepackage{hyperref}

\newcommand{\ket}[1]{\ensuremath{\left|#1\right\rangle}}
\newcommand{\Cre}[0]{\ensuremath{\hat{a}^{\dagger}}}
\newcommand{\Ann}[0]{\ensuremath{\hat{a}}}
\newcommand{\etal}{\textit{et al.\ }}

\begin{document}
\title{Dephasing of entangled atoms as an improved test of quantum gravity}

\author{Mark S\ Everitt, Martin L\ Jones and Benjamin T\ H\ Varcoe}
\affiliation{School of Physics and Astronomy,
  University of Leeds, Leeds, LS2 9JT,
  United Kingdom}

\date{\today.}

\begin{abstract}
In a recent article Wang \etal (Class. Quantum Grav. 23 (2006) L59), demonstrated that the phase of a particle fluctuates due to interactions with random deviations of a conformal gravitational field. Furthermore they demonstrated that atom interferometers are sensitive to these fluctuations and that sensitivity to Planck scale effects could be achieved with a sufficiently sensitive interferometer.  In this paper we demonstrate that a class of entangled states, the $N$-atom Greenberger-Horne-Zeilinger (GHZ) states, provide a better scaling than atom interferometers and that current experiments are capable of making a significant impact in this field. We outline an experiment which uses atomic beams of rubidium atoms excited to Rydberg states. The atoms undergo controlled collisions in high quality factor microwave resonators in a sequence that makes the resulting state highly sensitive to conformal field fluctuations. We show that a significant advance in sensitivity is possible.
\end{abstract}

\date{\today}
\maketitle

Entanglement is  unique in the quantum optics toolbox, and recent theory has demonstrated that an entangled matter interferometer has, for example, a significant advantage over a standard atom interferometer. Atom interferometers have been shown to be particularly sensitive to small gravitational fluctuations and have been put forward as the next generation of gravitational wave detector \cite{Lamine2002}. Like optical interferometers, the sensitivity of the interferometer grows with the square root of the number of particles involved \cite{Scully1993}. We propose using entanglement in a new class of interferometer, with which the sensitivity grows linearly with the number of particles, giving a potentially huge increase in sensitivity. This interferometer would be sensitive enough to probe the mesoscopic scale of gravity, yielding new information about Planck scale physics.

Extreme energy and length scales make the Planck scale difficult to reach directly. However over the last decade several steps have been taken towards phenomenological tests of quantum gravity. Power and Percival \cite{Power2000} showed that in the weak field and slow motion limit conformal fluctuations couple to massive particles and therefore atom interferometers should be able to detect the background field by making observations of small random relative shifts in phase. This approach was considered invalid as a treatment of quantum gravity as the conformal field is a constrained degree of freedom and is not a good candidate for quantisation. Wang \etal subsequently showed that, by a Hamiltonian constraint, vacuum fluctuations of gravitons should be paired with conformal fluctuations \cite{Wang2006}. Spectral decomposition of zero point gravitons leads to a prediction for paired conformal fluctuations, resulting in a similar result to Power and Percival but now accounting for the constrained nature of the conformal field.

In this letter we concentrate on the semiclassical domain described by Wang \etal \cite{Wang2006} where quantum gravitational effects are seen as random conformal fluctuations on otherwise flat Minkowski spacetime. We use the analogy that the dephasing seen in matter waves is like Brownian motion in that the origin of the fluctuations causing the dephasing is at the Planck scale but effects can be seen many of orders of magnitude away. The limit of these fluctuations is given as a cut-off length $l_\mathrm{cut}$, where $l_\mathrm{cut}$ is linked to the Planck scale by a factor $\lambda$, which is the subject of interest in \cite{Power2000,Wang2006}, by $l_\mathrm{cut}=\lambda l_\mathrm{Planck}$. $\lambda$ is a background dependent parameter, which should have a unique prediction from each theory of quantum gravity and it is expected to be in the $10^2-10^6$ range \cite{Bingham2005}. Following the analogy to Brownian motion, $\lambda$ is like a mean free path; larger values of lambda equate to less frequent kicks to the phase of massive particles and a lower rate of dephasing.

Wang \etal arrived at the linkage between dephasing of an atom \cite{Wang2006}
\begin{equation}
\label{dephase-lambda}
\lambda = \left(\frac{8c^4\tau_0\sqrt{2\pi^5}}{9\hbar^2}\cdot\frac{M^2T}{p}\right)^{1/3}\,,
\end{equation}
where $\tau_0$ is the Planck time, $M$ is the mass of the matter wave, $T$ is the time the matter wave is allowed to travel before recombination and $p$ is the parameter which describes dephasing. This equation is appealing because it provides a link between the dephasing of an atom and the background parameter $\lambda$, connecting each theory with a characteristic dephasing rate. Whatever this rate may be, it is likely to be swamped by dephasing caused by other sources in an experiment. Thus by surpassing a dephasing rate characteristic of a particular theory of quantum gravity it is ruled out. One can imagine a spectrum of theories, with an experiment exploiting this formula removing theories as it increases the minimum value of $\lambda$.

 The parameter $p$ is related to dephasing in the density matrix of an atom (a two state system) by
\begin{equation}
\label{dephasing}
\left(\begin{array}{cc}|\alpha|^2 & \alpha\beta^\ast \\\alpha^\ast\beta & |\beta|^2\end{array}\right)
\mapsto
\left(\begin{array}{cc}|\alpha|^2 & 0 \\0 & |\beta|^2\end{array}\right)
+ (1-p)\left(\begin{array}{cc}0 & \alpha\beta^\ast \\\alpha^\ast\beta & 0\end{array}\right)
\end{equation}
so that when $p=1$ the atom is entirely dephased and when it is 0 the atom is in a pure state.
For atom interferometry this is all that is needed. The loss of visibility in the interference fringes provides a lower bound on $\lambda$ by (\ref{dephase-lambda}). Atom interferometry experiments \cite{Peters1997} currently put a lower bound on $\lambda$ of 7600 \cite{Wang2006}. Unfortunately the prospects for atom interferometry experiments improving on this are limited as the sensitivity scales unfavorably. Equation (\ref{dephase-lambda}) shows that for every three orders of magnitude gain in sensitivity, only a single order of magnitude in $\lambda$ is achieved. It should be noted that atom interferometry uses a single atom, effectively referenced to itself. As an atom travels over one arm of the interferometer it will see a slightly different proper time compared with another atom following, and thus evolve in phase accordingly, due to conformal fluctuations. One atom split over two arms allows these small shifts in proper time to be measured after averaging over many atoms. Many atoms in the atom interferometer merely repeat the same experiment in parallel, allowing fringes to be seen and dephasing measured by loss of visibility.

Entanglement has been shown to be a sensitive tool for measuring dephasing and decoherence \cite{Vedral2008}. This is a problem in the field of quantum information, in which some entangled states which are particularly useful for quantum algorithms are unfortunately very sensitive to decoherence through dephasing. It is therefore interesting to consider the impact of quantum gravity induced dephasing on the correlations between measurements of entangled atoms.
Following Sung Jang \etal \cite{Jang2006}, the expectation of the measurement of an $N$-atom Greenberger-Horne-Zeilinger (GHZ) state that has been subjected to $p$ dephasing on each atom is
\begin{equation}
\label{GHZ-expect}
\left\langle B_1B_2\ldots B_N\right\rangle = \phantom+\frac{1+(-1)^N}{2}\left(\prod_{i=1}^N\cos\left(\theta_i\right)\right)
+ (1-p)^N\cos\left(\sum_{i=1}^N\phi_i\right)\left(\prod_{i=1}^N \sin\left(\theta_i\right)\right)\,,
\end{equation}
in an arbitrary basis $B_i$ for each atom $i=1,N$. The basis describes the orientation of measurement on the Bloch sphere using the angles $\theta$ and $\phi$,
\begin{equation}
B_i = \left[\sigma_x\cos\left(\phi_i\right)+\sigma_y\sin\left(\phi_i\right)\right]\sin\left(\theta_i\right)+\sigma_z\cos\left(\theta_i\right)\,,
\end{equation}
where $\sigma_\alpha$ are the Pauli operators. We choose $\theta = \pi/4$ and $\phi = 0$ to reduce (\ref{GHZ-expect}) to measurements only in $\sigma_x$. 
Physically this would be achieved by applying a resonant $\pi/2$ rotation in the Bloch sphere of an atom and then measuring it, thereby enabling us to detect one of the states $\ket{\pm}=(\ket{e}\pm\ket{g})/\sqrt{2}$. 
Under this condition (\ref{GHZ-expect}) reduces to
\begin{equation}
\langle \sigma_x^{\otimes N}\rangle = (1-p)^N\,.
\end{equation}
This can be substituted directly into (\ref{dephase-lambda}) to yield
\begin{equation}
\label{dephase-lambda-GHZ}
\lambda = \left(\frac{8c^4\tau_0\sqrt{2\pi^5}}{9\hbar^2}\cdot\frac{M^2T}{1-\langle \sigma_x^{\otimes N}\rangle^{1/N}}\right)^{1/3}\,.
\end{equation}
This is the central result of this letter. In the limit of $\langle \sigma_x^{\otimes N}\rangle \approx 1$ (\ref{dephase-lambda-GHZ}) reduces to
\begin{equation}
\lambda \approx \left(\frac{8c^4\tau_0\sqrt{2\pi^5}}{9\hbar^2}\cdot\frac{NM^2T}{1-\langle \sigma_x^{\otimes N}\rangle}\right)^{1/3}\,,
\end{equation}
which corresponds to the sensitivity of a typical experiment. This demonstrates an improvement with respect to atom interferometry of $N^{1/3}$. Unlike atom interferometers, this uses the electronic states of the atoms comprising an $N$-atom GHZ state to measure dephasing. This should be interpreted as atoms referenced against each other, rather than themselves. The phase evolution of an electronic state is of course still sensitive to fluctuations in proper time for each atom, and this can be used to measure dephasing of the atoms in the GHZ state.

We now look at entanglement as a measurement tool and determine the outcome of an experiment. 
The GHZ state is a pure state in which all parties are 0 or all parties are 1 in equal superposition. 
For example an $N=3$ GHZ state is $(\ket{000}+\ket{111})/\sqrt{2}$. 
This means that the decoherence of only one particle collapses the entire state leaving no entanglement regardless of the number of particles. These states are used for tests of non-locality.

The fragility of the GHZ class of states distinguishes it from other types of entangled state. The \textit{degree of entanglement} for any GHZ state of length $N$ is always 1, which is in contrast to the Werner state or `W-state' which has a degree of entanglement proportional to $\log(N)$. 
The W-state is simply an equal superposition of all states which contain $N-1$ zeros and a one (e.g. for $N=3$ the W state is $(\ket{100}+\ket{010}+\ket{001})/\sqrt{3}$). 
This kind of entanglement often occurs in systems with a large number of particles such as Bose-Einstein condensates. 
The sensitivity to dephasing grows with \textit{connectivity} as discussed by Vedral \cite{Vedral2008}. 
The connectivity of a GHZ state of $N$ atoms is $N$, whereas the connectivity of a W state is always 2 \cite{Vedral2008}. 
The W-state therefore retains the sensitivity equivalent to a pair of entangled particles and the advantage of entangling a large number of particles is lost.
These two states are opposite extremes of entanglement, but they make the point that one needs to be careful in designing the experiment to create the right type of entanglement. These two types of state have been shown to be fundamentally different \cite{Dur2000}, so it is important to define the type of entanglement that would be created in a specific experiment and show that it has the required properties.

We now give an example extensible system for producing GHZ states with atoms. Note that any system that produces an $N$-atom GHZ state may be suitable, and we give this example system for completeness.
In the experiment that we are proposing an entangled state is constructed between a number of independent atoms.
The entanglement is created using the collisional phase gate by Zheng and Guo \cite{Zheng2000} that was demonstrated by Haroche \etal \cite{Osnaghi2001}. 
\begin{figure}
\centerline{\includegraphics[width=3.1in]{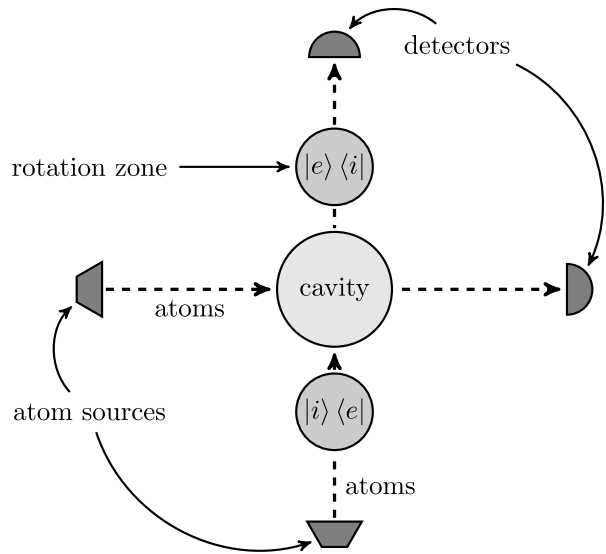}}
\caption[The Collisional Phase Gate]{A diagram demonstrating the collisional phase gate with two atoms as the qubits. The atom sources produce synchronized atoms with equal speed. The dashed arrows indicate the trajectory of individual atoms, the solid arrows are used to label components. This arrangement is a controlled phase gate on the atoms input in $(\alpha_1\ket{g}+\beta_1\ket{e})\otimes(\alpha_2\ket{g}+\beta_2\ket{e})$ with arbitrary $\alpha_{1,2}$ and $\beta_{1,2}$.\label{PhaseGateDiagram}}
\end{figure}
For this experiment we have in mind a ${}^{85}$Rb ladder of Rydberg states; $\ket{e}=$ 63P, $\ket{g}=$ 61D, and an auxiliary state $\ket{i}=$ 62P which are standard micromaser transitions \cite{Walther2006}.
Rydberg states are used commonly in the field of quantum optics for their relative longevity and large dipole moment \cite{Walther2006}. 
The atom is effectively a two state system, which is detuned from the cavity resonance by the amount $\Delta$. 
The Hamiltonian for this system is a two atom Tavis-Cummings Hamiltonian,
\begin{equation}
\frac{H_\mathrm{TC}}{\hbar} = \sum_i\left[\frac{\Delta\sigma^{(i)}_z}{2}+ig\left(\Cre\sigma^{(i)}_--\Ann\sigma^{(i)}_+\right)\right]\,,
\end{equation}
where the sum runs over the atoms in the interaction, $g$ is the atom-cavity dipole coupling constant, $\Cre$ ($\Ann$) are the bosonic creation (annihilation) operators acting on the single mode cavity field, $\sigma_+$ ($\sigma_-$) are the atomic raising (lowering) operators and $\sigma_z$ is the standard Pauli operator. This Hamiltonian is used to describe the evolution of a system with a single mode of the electromagnetic field coupling to a number of atoms simultaneously. In the situation with large detuning and two atoms, the system is reduced to an effective two-state system as the field is adiabatically eliminated. In this system an auxiliary state \ket{i} is sometimes used which is so far detuned from resonance that it may be assumed that it does not interact with the field at all. In this event the Hamiltonian reduces to a detuned Jaynes-Cummings model for the atom not in the auxiliary state. The detuned Jaynes Cummings interaction is given simply as (with the atoms in either order and zero photons in the field)
\begin{equation}
\begin{split}
\ket{g,i} &\mapsto \ket{g,i}\\
\ket{e,i} &\mapsto \mathrm{e}^{-i\gamma t}\ket{e,i}\,,
\end{split}
\end{equation}
where $\gamma$ is the effective coupling constant given by $\gamma=g^2/\Delta$. This interaction requires that the detuning is much larger than the coupling constant $\Delta\gg g$ to eliminate the field as a degree of freedom. The only two cases that they needed to consider for two atom interactions are the trivial case of two ground state atoms entering the cavity, which do not evolve, and the case of an excited atom and a ground state atom which interact by virtual excitation of the field. The case of two excited atoms entering is replaced by the case of one atom in the excited state and one in the auxiliary state. The reason and implementation of this omission will shortly become clear. For this case the Hamiltonian reduces to an effective Hamiltonian in the detuned limit
\begin{equation}
\frac{H_\mathrm{eff}}{\hbar}=\gamma\sum_{i=1,2}\left[\sigma^{(i)}_+\sigma^{(i)}_-\Ann\Cre-\sigma^{(i)}_-\sigma^{(i)}_+\Cre\Ann \right]
 +\gamma\sigma^{(1)}_+\sigma^{(2)}_-+\gamma\sigma^{(2)}_+\sigma^{(1)}_-
\end{equation}
For a cavity with zero photons this Hamiltonian yields the \ket{g,g} and \ket{e,g} evolutions
\begin{equation}
\begin{split}
\label{interaction}
&\ket{g,g} \mapsto \ket{g,g}\\
&\ket{e,g} \mapsto \mathrm{e}^{-i\gamma t}\left[\cos\left(\gamma t\right)\ket{e_1g_2} -i\sin\left(\gamma t\right)\ket{g,e}\right]
\end{split}
\end{equation}
In the lab a low thermal photon number ($\langle \Cre\Ann\rangle\approx 0$) is achieved by cooling the cavity following the same procedure as the micromaser experiment \cite{Walther2006}. By choosing the interaction time in (\ref{interaction}) to be $t=\pi/(4\gamma)$ with one atom in \ket{e} and the other in \ket{g}, an Einstein-Podolski-Rosen pair is produced. With the rotation of one atom to flip its state this is a GHZ state. Rotation of the atoms is achieved by passing the atoms through the side of a microwave waveguide. In order to produce atoms at the right time and velocity, the system uses atoms produced on demand \cite{Brattke2001}. Alternatively low Q cavities pumped with microwaves can be used. This allows production of atoms in both states \ket{e} and \ket{g} (although \ket{g} may be produced directly at the cost of more lasers and optics equipment), as well as the final rotation of states into the $\sigma_x$ basis for measurement.

The experiment outlined is capable of producing only $N=2$ GHZ states, and thus is sensitive to $(1-p)^2$. Extensions to this system allow for larger GHZ states to be prepared. GHZ states with $N$ atoms may be prepared by interacting one atom with the other atoms in turn in a similar scheme to that for 2 atoms. For the more general case, atoms are prepared in \ket{+} states initially. We also invoke the auxiliary level \ket{i}, as we are implementing the true phase gate as outlined by Zheng and Guo \cite{Zheng2000}. As stated above this state does not interact with the field as it is not resonant.

\begin{figure}[ht]
\centerline{\includegraphics[width=3in]{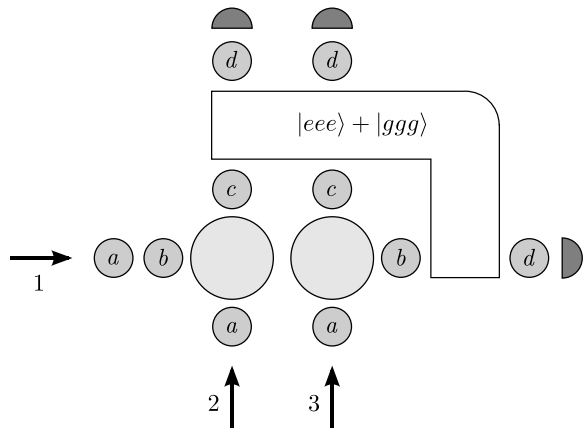}}
\caption[Preparing a five party GHZ state]{This example produces a three party GHZ state of atoms. Atoms approach as indicated by arrows to coincide in each cavity. Initially all atoms are in state \ket{e}. The rotation zones labelled \textit{a} are $\theta=3\pi/2$ pulses and $c$ are $\theta=\pi/2$ pulses. Those zones labelled \textit{b} act on the first atom (traveling horizontally) to switch \ket{e} components of the state to \ket{i}, the auxiliary state, and vice versa. The enclosed region is where the GHZ state exists as it travels for time $T$ related to velocity of atoms $v$, which is not to scale.\label{gravity_5GHZ}}
\end{figure}

For the general system rotation zones rotate every atom into a \ket{+} state before interaction. $N-1$ cavities are placed in a line, and a special rotation zone is placed at either end. These rotation zones transfer \ket{e} to \ket{i} and vice versa. One of the atoms passes through every cavity and these rotation zones, interacting with one of the other atoms in each cavity with interaction time $t=\pi/\gamma$. A $\pi$ pulse is then applied to this atom to flip its state, and a $\pi/2$ pulse to the other atoms. The GHZ state is now prepared and allowed to fly as long as possible for gravitational dephasing to take effect. Finally a $\pi/2$ pulse is applied to every atom for measurement in the $\sigma_x$ basis. A diagrammatic example based on a current experimental design \cite{Blythe2006}, is given in figure \ref{gravity_5GHZ} for a 5 atom GHZ state. The experiment is run multiple times to develop the expectation value $\langle \sigma_x^{\otimes N}\rangle$ for substitution into (\ref{dephase-lambda-GHZ}), placing a new lower bound on $\lambda$.

In summary we have shown that $N$-atom Greenberger-Horne-Zeilinger entangled states provide an avenue of increasing the sensitivity of tests of quantum gravity by a factor of $N^{1/3}$ over single atom interferometry; a significant gain in contrast with the difficulty of increasing the mass of a particle to interfere in an interferometer. We have also shown how this state may be produced in the laboratory and that the effect of gravitational dephasing can be measured using current apparatus. This is therefore within current technological bounds and can be accessed by leveraging existing quantum optics technology.

We would like to thank Charles Wang and Pieter Kok for useful discussions. This work was supported by the United Kingdom EPSRC 
through an Advanced Fellowship GR/T02331/01,
Scholarship GR/T02324/01 and
Grant GR/S21892/0.


\end{document}